\documentclass[aps,prl,reprint,nofootinbib]{revtex4-1}
\usepackage{savesym}
\usepackage{graphicx}
\expandafter\let\csname equation*\endcsname\relax
  \expandafter\let\csname endequation*\endcsname\relax 
\usepackage{subfig}
\usepackage{amsmath}
\usepackage{amssymb}
\usepackage{verbatim}
\usepackage[yyyymmdd,hhmmss]{datetime}
\usepackage{array}
\usepackage{times}
\usepackage[total={17.8cm,24.0cm},centering]{geometry} 
\usepackage{color}

\newcommand{\beq}{\begin{equation}}
\newcommand{\eeq}{\end{equation}}

\newcommand\R {{r_I}}
\newcommand\ji {{J}}
\newcommand\ei {{\gamma}}
\newcommand\U {{U^r}}
\newcommand\dd {\partial}
\newcommand\D {{\cal D}}

\begin{document}

\title{Inspirals from the innermost stable circular orbit of Kerr black holes: \\ Exact solutions and universal radial flow}

\author{Andrew Mummery}
\email{andrew.mummery@physics.ox.ac.uk}
\author{Steven Balbus}
\affiliation{Oxford Astrophysics, Denys Wilkinson Building, Keble Road, Oxford, OX1 3RH, United Kingdom}

\begin{abstract}
We present exact solutions of test particle orbits spiralling inward from the innermost stable circular orbit (ISCO) of a Kerr black hole.    Our results are valid for any allowed value of the angular momentum $a$-parameter of the Kerr metric.   These solutions are of considerable physical interest.   In particular, the radial 4-velocity of these orbits is both remarkably simple and, with the radial coordinate scaled by its ISCO value, universal in form, otherwise completely independent of the black hole spin. 
\end{abstract}

\maketitle

\section{1.  Introduction}

Classes of exact orbital solutions in the full Kerr geometry are a known, but under-utilised commodity.   Examples include pure circular orbits, radial plunges \cite{rev1}, so-called zoom-whirl orbits, and homoclinic orbits, which separate long-lived bound and plunging states \cite{LP1} (see \cite{LP2} for a useful `periodic table' of different black hole orbits).  The study of relativistic test-particle orbits characterised by the energy and angular momentum of a circular orbit, but which are not moving on that orbit, is not new \cite{Darwin}, but has yet to be astrophysically fully exploited.  In this {\it Letter,} we analyse an important sub-class of these orbits, and present exact Kerr orbital solutions in a parameter regime of direct physical interest to black hole accretion. 
While simple in mathematical form, these solutions exhibit revealing features which are important for understanding the accretion process, but have not been discussed before.

One of the most salient features of orbits in Kerr spacetimes is the existence of an innermost stable circular orbit, or ISCO.   Exterior to the radial ISCO coordinate $r=r_I$, circular orbits are stable and approach their Keplerian $1/\sqrt{r}$ {velocity} behaviour on scales large compared to the horizon radius.    Interior to $r=r_I$, the angular momentum of a circular orbit increases inward, which means that the orbits are unstable: a tiny perturbation from circular motion will eventually acquire a significant inward radial velocity, even while formally conserving its angular momentum and energy.   

Orbits interior to the ISCO are of astrophysical interest because of their direct relevance to black hole accretion  theory (e.g., \cite{SS, NT, PT}).   In particular, the question of whether, and if so under what conditions, there can be significant X-ray emission or other observational signatures from matter flowing inward from the ISCO is an active area of current research \cite{Wilkins,Fabian}.     This problem is generally approached via numerical techniques (e.g., \cite{Wilkins, Potter}), as the assumptions of the classical analytic ``viscous''  solutions of black hole accretion theory completely breakdown at, and within, the ISCO.  
The classical viscous disc solution for $U^r$ diverges at the formal ISCO radius \cite{NT, PT}, and without a more fundamental understanding of the inflow dynamics, it is not even clear how to frame the underlying equations!  
With a few notable exceptions \cite{Rey, Wilkins, Fabian} virtually all existing accretion models are artificially cut-off at the ISCO. 

In this work we show that there is an overlooked but dramatic simplification of this problem, which provides a clear path forward.   The implicit averaging procedure associated with viscous (more accurately, turbulent) disc theory no longer makes sense when the orbits are not circular, but plunging \cite{Rey}.    The need to shed angular momentum vanishes.   {Instead, in the Kerr geometry angular momentum is simply advected inward with the fluid elements, and thus remains approximately constant and independent of position.} 
What is new here is an explicit solution for the Kerr radial 4-velocity, which determines the surface density directly from mass conservation, and is
is both simple and universal in form.   This is a key result of this paper.   With $U^r$ known, we present exact, closed analytic solutions to the general problem of a test particle starting at $r=r_I$ at time $t=-\infty$, and thereafter inspiraling toward the origin.  The  orbital shape (radial coordinate as a function of azimuthal angle $\phi$) is determined entirely in terms of elementary functions.     For the Schwarzschild geometry, this orbital shape is exceptionally simple (see eq. [\ref{pisco}]).    Despite making an appearance in Chandrasekhar's classic text \cite{Chand}, it seems to have remained dormant and largely unreferenced in the astrophysical literature.   The orbital solutions for more general Kerr geometries, also of astrophysical interest but which seem not to have been discussed in the literature,  are best expressed in parameterised $r(\psi), \phi(\psi)$ form, in a manner analogous to the classical Friedman matter-dominated cosmologies.

\section {2.  Preliminary analysis} 
Throughout this work we use geometric units in which both $c$, the speed of light, and $G$, the Newtonian gravitational constant, are set equal to unity.    In coordinates $x^\mu$, the invariant line element  
is ${\rm d}\tau^2 = - g_{\mu\nu} {\rm d}x^\mu {\rm d}x^\nu$,
where $g_{\mu\nu}$ is the usual covariant metric tensor with spacetime indices $\mu, \nu$.    (We use the signature convention 
 $-+++$.) The coordinates are standard $(t, r, \theta, \phi)$ Boyer-Lindquist, where $t$ is time as measured at infinity, and the other symbols have their usual quasi-spherical interpretation.    We shall work exclusively in the Kerr midplane $\theta =\pi/2$. For black hole mass $M$ and angular momentum $a$ (both having dimensions of length in our choice of units), the non-vanishing $g_{\mu\nu}$ required for our calculation are presented here for convenience (e.g. \cite{HEL}):
\begin{align}
g_{00} &= -1 +2M/r, \quad  g_{0\phi} = g_{\phi0} = -2Ma/r,  \nonumber \\
g_{\phi\phi} &= r^2+a^2 +2Ma^2/r, \quad  g_{rr} = r^2/(r^2 - 2Mr +a^2). 
\end{align}

The 4-velocity vectors are denoted by $U^\mu={\rm d}x^\mu/{\rm d}\tau$.  {Their covariant counterparts $U_\mu$, in particular $U_0$ and $U_\phi$, have a significance as conserved quantities and are discussed further below.   Test particles orbits which spiral inwards, starting at a distant time $t=-\infty$ from a circular orbit at $r=r_I$, will preserve their energy and angular momentum. }  
General expressions for the circular angular momentum and energy at radius $r$  may be found in \cite{HEL}:  
\begin{align}
U_\phi &= {(Mr)^{1/2}\over \D}(1+a^2/r^2-2aM^{1/2}/r^{3/2}), \label{am1}\\
U_0 &= - {1\over \D} (1-2M/r +aM^{1/2}/r^{3/2}) \label{am2},
\end{align}
where 
\beq
\D^2=1 -3M/r +2aM^{1/2}/r^{3/2} .
\eeq
The circular orbits described by eqs. (\ref{am1}) and (\ref{am2}) are not stable at all radii, and it may be shown (e.g., \cite{HEL}) that these orbits are stable {\it only} when the following condition is satisfied: 
\beq
\left( 1 - {6M\over r} -{3a^2\over r^2} +{8aM^{1/2}\over r^{3/2}}\right)  > 0, 
\eeq 
{which corresponds to $\dd_r U_\phi > 0$.  } 
The location of marginal stability $r=r_I$ corresponds to this expression vanishing:
\begin{align}\label{a}
r_I^2  &=    6Mr_I-8a\sqrt{Mr_I} +3a^2 \nonumber\\
& = {2Mr_I\over 3}+{16Mr_I\over 3}\left( 1 - {3a\over 4\sqrt{Mr_I}}\right)^2 .
\end{align}
The form of the second equality will be especially convenient in what follows below.
 
We label the constants of motion $J = U_\phi(r_I),  \gamma = - U_0(r_I)$, and use equation (\ref{a}) to substitute for $r_I^2$. The resulting numerators and denominators factor cleanly, leading to an additional simplification:
\begin{align}\label{J}
J &= 2\sqrt{3}M\left( 1 - {2a\over 3\sqrt{Mr_I} }\right) , \\
\gamma &= {4\over3}\sqrt{3}\left(M\over r_I\right)^{1/2} \left( 1 -{3a\over 4\sqrt{Mr_I}}\right) \label{gam}  = \left(1- {2M\over 3r_I}\right)^{1/2},
\end{align}
where in the second $\gamma$ equality we have made use of equation (\ref{a}).    Notice that $\gamma$ is independent of $a$, apart from the simple implicit $r_I$ dependence.

\section  {3.\ Orbital Solutions}
\subsection {3a. Radial velocity of the ISCO inspiral }

As noted earlier, general relativistic dynamics allows for radial motion in orbits whose angular momentum and energy values correspond to an exactly circular orbit.  
While the governing equation is easily stated
\beq
g_{rr} (U^r)^2 +U^0U_0 +U^\phi U_\phi =-1, 
\eeq
its direct solution is generally a matter of some algebraic complexity.   
Expressing all non-radial 4-velocities in terms of $J$ and $\gamma$, and multiplying through by $1/g_{rr}$, we have
\begin{multline}\label{full}
 (\U)^2 + {\ji\over r^2} \left( {2Ma\ei \over r} + \left(1 - {2M \over r}\right) \ji \right) \\ - {\ei\over r^2} \left[\left(r^2 + a^2 + {2Ma^2 \over r}\right) \ei - {2Ma\ji \over r}\right]   = -1-{a^2\over r^2} +{2M\over r}.
\end{multline}
Equation (\ref{full}) is of the form $(\U)^2 + V_{\rm eff}(r) = 0$,  where $V_{\rm eff}$ is a cubic in $1/r$ which may always be factored:
\beq
 V_{\rm eff}(r) = - V_0 \left({r_1\over r} - 1\right)\left({r_2\over r}- 1\right)\left({r_3\over r}-1\right),
\eeq 
where $r_1, r_2$ and $r_3$ are the general (possibly complex) roots of $V_{\rm eff}$.   {($V_{\rm eff}$ is, up to an irrelevant  factor of $2$, the usual effective potential.)}

For an arbitrary circular orbit of radius $r=r_c$,  both $V_{\rm eff}(r_c) = 0$ and $\partial_r V_{\rm eff}(r_c) = 0$, and $r_c$ will thus be a double root of the polynomial.    For the particular case of a {\it marginally stable} circular orbit, there is an additional condition, $\partial^2_r V_{\rm eff}(r_c) = 0$.  Thus, $r_I$ must be a {\it triple} root of $\U$.   The normalisation constant $V_0$ may be found by going back to equation (\ref{full}) and taking the formal limit $r\rightarrow \infty$. 
We find
\beq
V_0
= 1- \gamma^2 =   {2M \over 3 r_I} ,
\eeq   
which leads directly to our final equation for 
$U^r$:
\beq\label{flow}
\U \equiv  {{\rm d}r\over {\rm d}\tau} = - \sqrt{2M\over 3\R} \left({\R \over r} - 1\right)^{3/2}.
\eeq
This may also be verified by a (considerably more lengthy!) direct computation.  
Note  the universality of this remarkable result:  there is no $a$-dependence in this expression other than implicitly through $r_I$.  
Every Kerr orbit inspiraling from an ISCO is self-similar in its radial motion.   As  expected, no radial velocity solutions exist for $r>r_I$.     Despite its generality, simplicity and importance, equation [\ref{flow}] does not appear to have been recognised before in the literature.   

The azimuthal component of the intra-ISCO 4-velocity is also simple:
\beq\label{uphi}
U^\phi = - g^{\phi 0} \ei + g^{\phi\phi} \ji = {2M\ei a + \ji(r -2M) \over r(r^2 - 2Mr + a^2)} .
\eeq

\subsection{3b.  Schwarzschild orbits}

We begin with the Schwarzschild limit $a= 0$.   Then  $r_I=6M$, $J=2\sqrt{3}M$,  and $U^\phi=2\sqrt{3}M/r^2$.  Defining $x=r/r_I$, we find
\beq
{{\rm d}\phi\over {\rm d}x} = 6M{U^\phi \over U^r} = - \sqrt{3} {1\over x^2} \left({1\over x} - 1\right)^{-3/2},
\eeq
which immediately integrates to
\beq\label{pisco}
x = {r\over 6M} = {1\over 1+12/\phi^2},
\eeq
with the convention that $\phi$ increases from $-\infty$ to $0$ as $x$ goes from $1$ to $0$.    {\em This is an {\it exact} orbital solution for the standard Schwarzschild metric, which is both non-circular and non-radial, extending from $r_I$ to $r=0$.}  
The reader may verify this by direct substitution into the exact Schwarzschild orbit equation for $u=1/r$:
\beq
{{\rm d}^2u\over {\rm d}\phi^2} + u = {M\over J^2} +3Mu^2.
\eeq

Equation (\ref{flow}) is also easily integrated.   This result, moreover, holds for {\em any}  Kerr ISCO-inspiral orbit, not just for those in the Schwarzschild geometry.    With $x=r/r_I$, we find 
\beq
\tau = \sqrt{3r_I^3\over 2M} \left( 3\sin^{-1}\sqrt{x} +(x-3)\sqrt{x\over 1-x}\right)
\eeq 
a universal relationship between proper time $\tau$ and coordinate $r$.    It may also be written in parametric form, reminiscent of closed Friedmann cosmologies:
\begin{align}
x &= {1\over 2}(1-\cos\psi)=\sin^2(\psi/2), \\
\tau &= \sqrt{3r_I^3\over 2M}\left[ {1\over 2}(3\psi - \sin\psi)-2\tan\left(\psi\over2\right)\right] ,
\end{align}
with $\psi$ running from $\pi$ to $0$.

\subsection {3c.  General solution}

\begin{figure}
 \includegraphics[width=.4\textwidth]{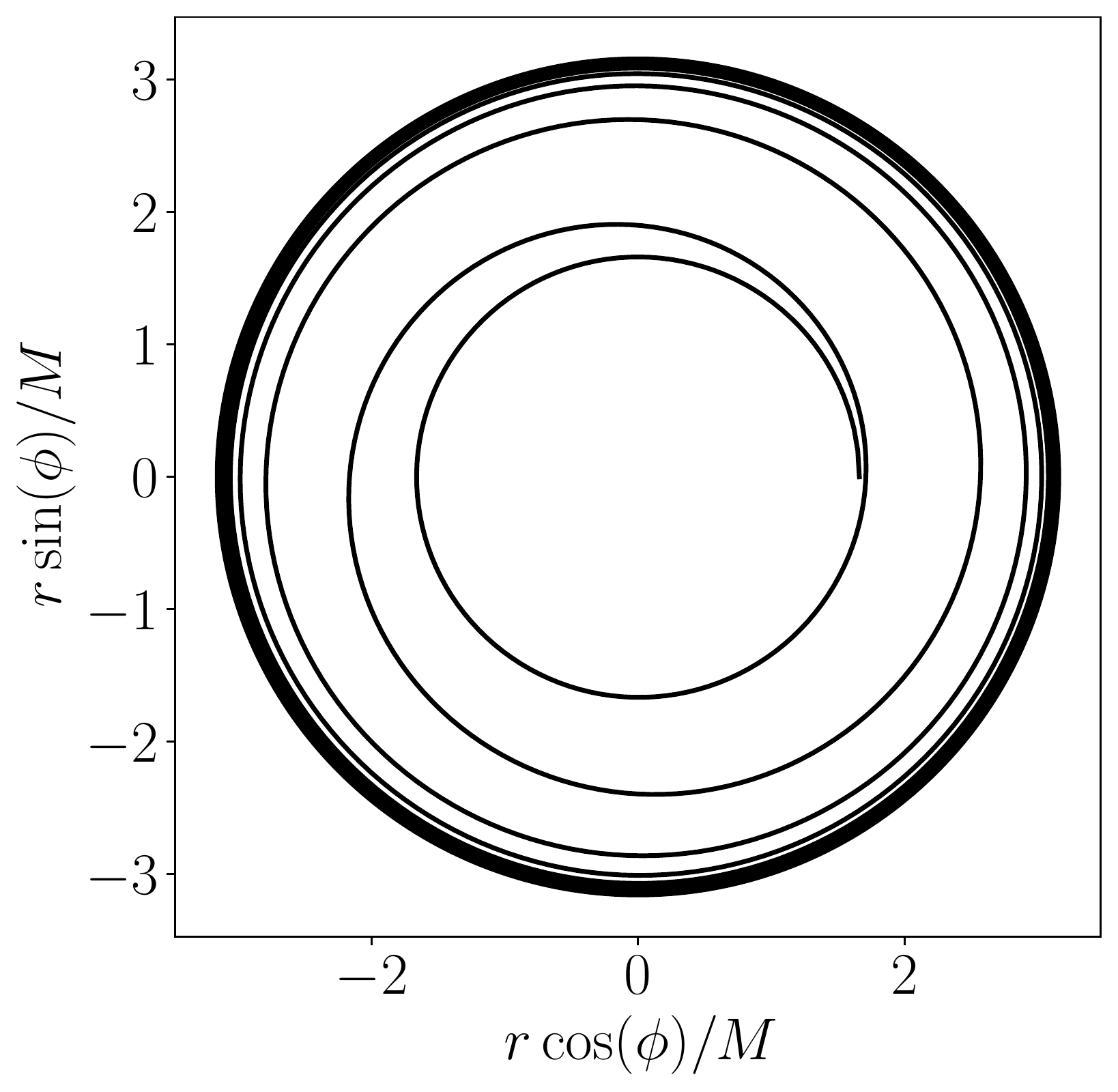} 
 \caption{Inspiral (eq.  [\ref{phisol}]) of a test particle from ISCO ($r \simeq 3.158M$) to event horizon ($r_+ \simeq 1.661M$) of a Kerr black hole, $a/M = +0.75$.  } 
 \label{ParSol}
\end{figure}

The general parameterised azimuthal solution $\phi(\psi)$ is considerably  more complicated.    Start with 
\beq
{{ \rm d} \tau\over {\rm d}\psi} =  -\sqrt{3r_I^3\over 2M}\sin^2(\psi/2) \tan^2(\psi/2) ,
\eeq
whence
\beq
{{\rm d}\phi\over {\rm d}\psi} = - U^\phi  \sqrt{3r_I^3\over 2M}\sin^2(\psi/2) \tan^2(\psi/2).     
\eeq
With $U^\phi$ substituted from (\ref{uphi}) and $r= r_I\sin^2(\psi/2)$, this equation has the formal solution
\beq\label{phiint} 
\phi (\psi) = \sqrt{3r_I^3\over 2M}{1\over r^3_I} \left[ 2M(J-a\gamma)I_1-Jr_I I_2\right] ,
\eeq
where $I_1$ and $I_2$ may be written in the single compact form:
\beq
I_j =\int^\psi_\pi  {\tan^2(\psi'/2) \left(1+[1-j] \cos^2(\psi'/2)\right) {\rm d}\psi'\over \sin^4(\psi'/2) -2M\sin^2(\psi'/2)/r_I + a^2/r_I^2} .
\eeq
Both $I_1$ and $I_2$ have poles at the Kerr event horizons, $r_+$, $r_-$, at the $\psi$ values given by 
\beq\label{poles}
r_I\sin^2(\psi_\pm/2) = r_\pm = M\pm\sqrt{M^2-a^2} .
\eeq
\begin{figure}
\includegraphics[width=.45\textwidth]{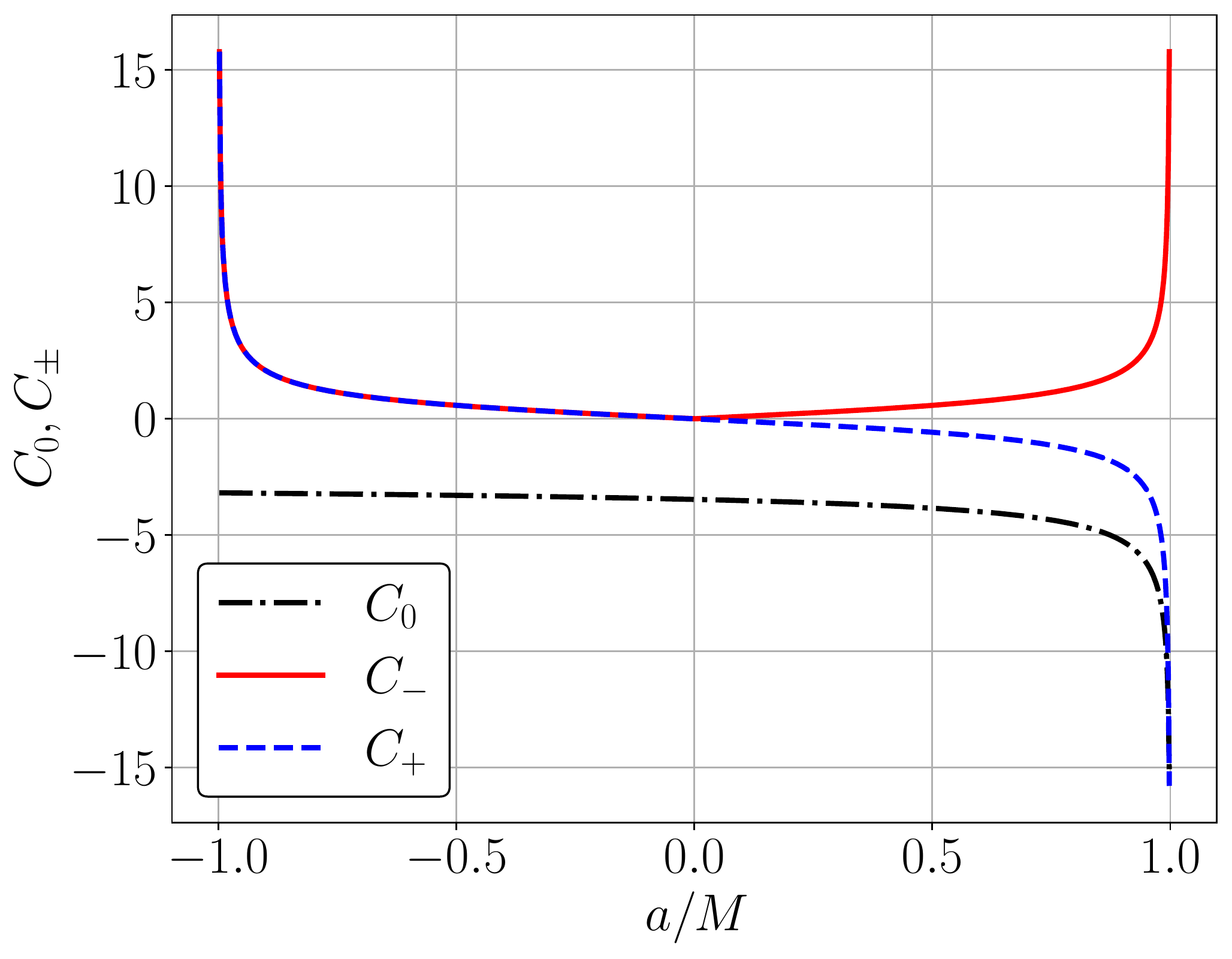} 
 \caption {Coefficients $C_0, C_+$ and $C_-$ appearing in the parametric $\phi(\psi)$  \ \ solution (eqs. \ref{c0}, \ref{cpm}),  as a function of \ \ \  $a/M$.   (Spin axis plotted for $-0.998 < a/M<0.998$.) } 
 \label{coeffs}
\end{figure}
$I_1$ and $I_2$ may be evaluated by the Weierstrass substitution
\beq
t= \tan(\psi'/2).
\eeq
We then find $I_1 = K_1 + K_2$, $I_2=K_2$, where 
\begin{align}
K_j &= \int {2t^{2j} \over \alpha t^4 - 2\beta t^2 + a^2/r_I^2} \, {\rm d}t, \\
\alpha &= 1 - 2M/r_I +a^2/r_I^2, \quad \beta = M/r_I -a^2/r_I^2 .
\end{align}
The two roots of the denominator of the $K$-integrals are
\beq
t^2_\pm = {\alpha^{-1}}\left(\beta \pm \sqrt{\beta^2-a^2/r_I^2}\right) ,
\eeq
which, from [\ref{poles}], are
\beq
t_\pm = \tan(\psi_\pm/2) = \sqrt{x_\pm \over 1 - x_\pm}, \quad x_\pm \equiv r_\pm/r_I . 
\eeq
After factoring the denominators, the $K$-integrals may be evaluated by partial fraction expansions.     Omitting the lengthy but straightforward details, the final result for $\phi$ is
\begin{multline}
\phi(\psi) = C_0 \tan(\psi/2)  + C_- \tanh^{-1}\left({\tan(\psi_-/2) \over \tan(\psi/2)}\right) \\
- C_+ \tanh^{-1}\left({\tan(\psi_+/2) \over \tan(\psi/2)}\right). 
\end{multline}
We have defined the constants 
\begin{align}
C_0 &= \sqrt{6r_I\over M}\ {2M(J-a\gamma) - r_IJ\over r^2_I -2Mr_I +a^2} \label{c0} ,\\
C_\pm &=  {t_\pm\over t_+^2 - t_-^2}\sqrt{6r_I\over M}\ {2M(J-a\gamma)(1+t_\pm^2) -r_IJt_\pm^2 \over r^2_I -2Mr_I +a^2}.  \label{cpm}
\end{align}
This can also be written explicitly as a function of radius:  
\begin{multline}\label{phisol}
\phi(r) = C_0\sqrt{r \over r_I - r}  + C_- \tanh^{-1}\left(\sqrt{{r_- \over r}{ (r - r_I) \over (r_--r_I)}}\right) \\
- C_+ \tanh^{-1}\left(\sqrt{{r_+ \over r}{ (r - r_I) \over (r_+ -r_I)}}\right). 
\end{multline}
(The arguments of the $\tanh^{-1}$ functions should be inverted for radial coordinates within the horizons $r < r_\pm$.) 
The $C$-coefficients are plotted as a function of black hole spin in fig. [\ref{coeffs}].  The Schwarzschild limit $a\rightarrow 0$ and $r_I=6M$, for which $C_\pm=0$, $C_0=-2\sqrt{3}$, is easily verified.   An example inspiral trajectory given by equation \ref{phisol} with $a/M = +0.75$ is shown in fig. [\ref{ParSol}]. These solutions have been verified against numerical integration of the geodesic equations.

\subsection{3d.  Extremal  spin limit}

The above solution for $\phi(r)$ is ill-defined in the maximal $a/M = \pm 1$ limit, as the two event horizons of the Kerr geometry coincide ($t_+ = t_-$), and the partial fraction approach used in solving equation (\ref{phiint}) must be revisited. For the $a = +M$ solution,
the ISCO and event horizons formally coincide in Boyer-Lindquist coordinates.   For $a = -M$ the orbit is more interesting, and we are able to solve exactly for $r(\phi)$.  We rewrite equation \ref{phiint} (with $\ji = 22\sqrt{3}M/9$, $\ei = 5\sqrt{3}/9$, $r_I = 9M$, $a = -M$ and $t = \tan[\psi/2]$) as
\beq
{\phi \over 9\sqrt{2}} =  \int{6t^2 \over (8t^2 - 1)^2} \, {\rm d}t -  \int{16t^4 \over (8t^2 - 1)^2} \, {\rm d}t ,
\eeq
which becomes 
\beq\label{38t}
{\phi \over 9\sqrt{2}} = {2t^3 \over 1 - 8t^2} .
\eeq
With $t^2=r/(r_I-r)$, we find for $\phi(r)$:
\beq\label{43}
\phi(r) = {2\sqrt{2} \over 3 M^{3/2} } {r^{3/2} \over (1 - r/M) \sqrt{1 - r/9M}} .
\eeq

Inverting equation (\ref{38t}) to solve for $t(\phi)$ (and thus $r(\phi)$) is interesting, as it highlights the solutions from all branches of the resulting cubic equation
\beq\label{cubic}
t^3 + {4 \phi \over 9 \sqrt{2}} t^2 - {\phi \over 18\sqrt{2}} = 0.
\eeq 
In fact, there are actually {\em four} branches of interest in this problem!  Physically this arises from frame dragging, which produces a multi-valued $\phi(r)$.  The three nominal roots of the cubic, $t_j$, may be written:
\begin{align}
t_j &= {2\sqrt{2} \phi \over 27} \left[2 \cos\left({1 \over 3}  \cos^{-1}\left[ {2187\over 128\phi^2} - 1\right]+{ 2\pi j \over 3} \right)  - 1 \right], \label{t1}
\end{align}
with $j = 0, 1, 2$.   The fourth branch $t_3$ may be formally identified with the ($j=0$) $t_0$ root, but one must change $\cos$ and $\cos^{-1}$ into $\cosh$ and $\cosh^{-1}$ respectively, as the argument of the $\cos^{-1}$ exceeds unity for $t_3$ orbits.   All four branches are needed, as we now discuss.

\begin{figure}
  \includegraphics[width=.4\textwidth]{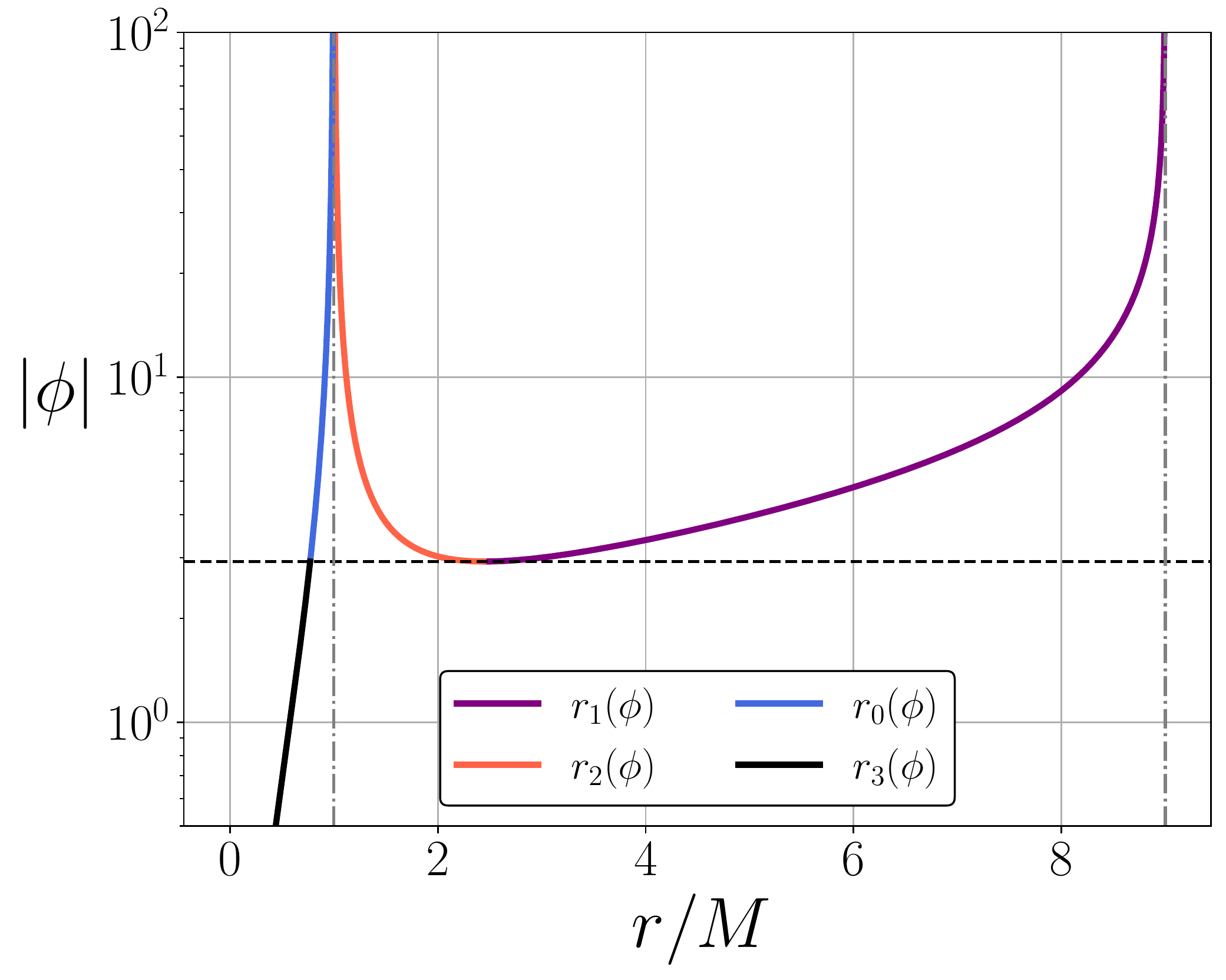} 
  \includegraphics[width=.4\textwidth]{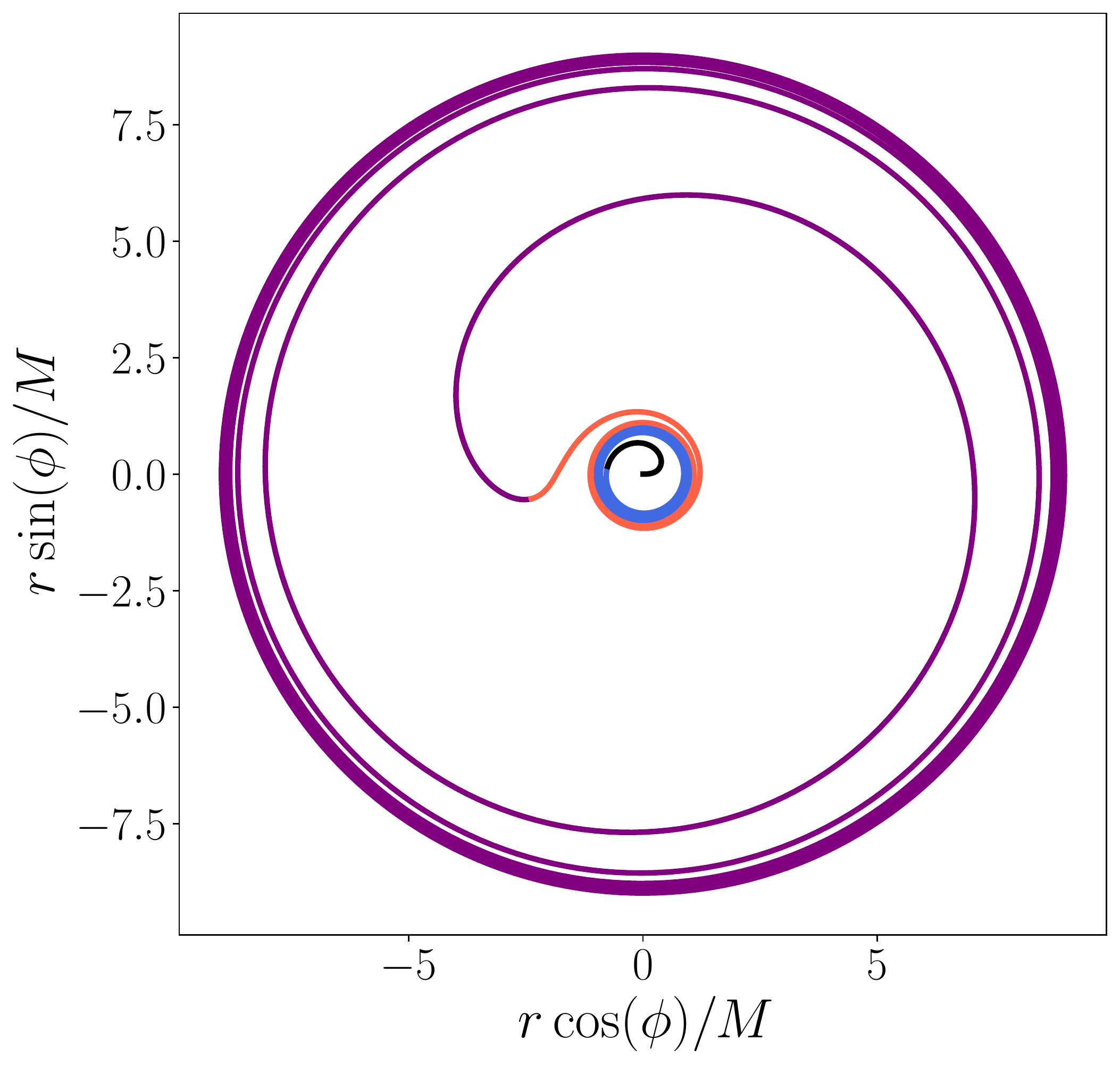}
  \caption{Inspiral of a test particle from ISCO ($r = 9M$) to singularity of an extremal Kerr black hole $a = -M$ (eq.\ [\ref{43}] {\it et seq.}).  Upper figure shows the magnitude of the $\phi$ coordinate over the infall, with the four $r_j(\phi)$ branches of the solution color-coded (see text). The value of the frame-dragging angle $\phi_\star$ (eq.\ [\ref{phisstar}]) is shown by the horizontal dashed line. The lower figure shows the inspiral in quasi-Cartesian coordinates with the same color-coding scheme.  Note the pronounced effect of counter rotating frame dragging upon the orbit.  } 
 \label{maximal_inspiral}
\end{figure}

The radial solutions for these roots are
 \beq
 r_j = {9M t_j^2 \over 1 + t_j^2}, 
 \eeq
with different roots corresponding to different ``legs'' of the orbit.    
The path of a test particle inspiralling from the retrograde ISCO of a maximally rotating Kerr black hole is as follows.   The particle starts at $r = 9M$, with $\phi = -\infty$.  It spirals inwards towards the event horizon, with increasing $\phi$, until it crosses the special radius $r_\star = 27M/11$ at an angle 
\beq\label{phisstar}
\phi_\star = -{27\sqrt{3} \over 16},
\eeq
whereupon frame-dragging bends the orbit backwards,  and the particle begins to corotate with the black hole (Fig. \ref{maximal_inspiral}).   Note that this location is exterior to the ergosphere $r_E = 2M$.   During this initial  phase the radial coordinate is given by the $r_1(\phi)$ solution (eq. \ref{t1}).  Beyond this point, the orbit transitions onto the $r_2(\phi)$ branch, with $\phi$ once again tending towards $-\infty$ as the particle approaches the event horizon $r_H = M$. Within the event horizon, the orbit is first described by the $r_0(\phi)$ expression, with $|\phi_\star| < \phi < \infty$, and then the final branch $r_3(\phi)$ for $0 < \phi < |\phi_\star|$ (Fig \ref{maximal_inspiral}).

\section {4. Discussion: accretion inside the ISCO }

The most important result of this {\em Letter} for the modelling of black hole accretion flows is also the simplest, namely equation (\ref{flow}) for $\U(r)$, the universal form of the radial 4-velocity for a test particle insprialling from the ISCO radius of a Kerr black hole while retaining its circular energy and angular momentum.   This is of great astrophysical interest because the accretion of a steady mass flow $\dot M$ in a thin disk is characterised by a constant value of the mass flux $2\pi \Sigma  r U^r $ (here, $\Sigma$ is the disk surface density).   Here, we have learned that within the ISCO, $U^r$ is generally known {\it a priori} and is extremely simple (at least for midplane orbits).  A stress term is not needed in this region, and may be ignored since now gravity alone is much more efficient at powering inward flow. By way of contrast, external to the ISCO,  to have any systematic radial velocity at all,  an enhanced stress is absolutely essential \cite{PT, NT}.    With this external stress remaining finite at the ISCO itself, the interior and exterior solutions both have vanishingly small velocities at $r_I$,  so that joining these two branches to form a global black hole accretion solution is a natural, and potentially very revealing, next step.

The energetics of the post-ISCO flow is also very important, and becomes much more accessible once the radial velocity is known {\it a priori}.   The prompt acceleration results in a rapid drop in $\Sigma$ and therefore the hot internal radiation field will escape more easily.   Offsetting this is a radial expansion producing global cooling. This radial expansion is to some extent offset by the vertical compression of the flow near the outer Kerr event horizon,  a process which acts to heat the flow.  An understanding of the interaction between these competing adiabatic effects, together with self-heating of the disc by its {\it own emission} will be needed to account for recent observations that appear to show an additional hot disk component, beyond the expectations of standard theory, in some sources \cite{Fabian}.   

A more precise  knowledge of $U^r$ is a very important aid for numerical accretion disk modellers.  With a simple form for $U^r={\rm d}r/{\rm d}\tau$, and detailed mathematical formulae for $r(\tau)$ and $r(\phi)$, these results promise to be of great practical utility for numerical code calibration \cite{Porth}. 

Finally, the imaging capabilities now available through the Event Horizon Telescope are on the verge of resolving the flow between the ISCO and event horizon itself \cite{Aki}. The dynamic properties of the flow in the intra-ISCO region may one day soon be imaged directly. The data from such a study   would no doubt provide interesting tests of the models derived here.

We conclude by noting that the mathematical methods used here also work well for other classes of noncircular orbits, which are defined by the energy and angular momentum of a non-ISCO circular orbit, and also for photon orbits.      The analytic results of these studies will be fully described in a more lengthy investigation to follow.   

\bigskip

{\it Acknowledgements:}      We thank R.\ Blandford, K. Clough, P. Feirrera, C. Reynolds and  J. Stone  for stimulating conversations and helpful advice. 
This work is partially supported by the Hintze Family Charitable Trust and STFC grant ST/S000488/1.

\label{lastpage}
\end{document}